\input{aipcheck}

\documentclass[
    ,final            
  ]
  {aipproc}

\layoutstyle{6x9}

\newcommand{\beq}{\begin{equation}}
\newcommand{\eeq}{\end{equation}}

\begin{document}

\title{Indirect Dark Matter Searches in the
Light of ATIC, FERMI, EGRET and PAMELA
}
\date{\today}

\pacs{95.35.+d, 95.85.Pw, 98.65.-r, 98.70.Sa}

\classification{95.35.+d, 95.85.Pw, 98.65.-r, 98.70.Sa}
\keywords      {Dark Matter, Galaxy, Cosmic Rays}

\author{W. de Boer}{
  address={Institut f\"{u}r Experimentelle Kernphysik\\
KIT, D-76131 Karlsruhe, Germany} }


\begin{abstract}
 Recently, new data on antiprotons and positrons from PAMELA, e$^-$+e$^+$ spectra from ATIC, FERMI and
 HESS up to TeV energies all indicate deviations from expectations, which has caused an interesting mix
 of new explanations, ranging from background, standard astrophysical sources to signals from dark matter
 (DM) annihilation. Unfortunately, the excess in positrons is not matched with obvious excesses in antiprotons or
 gamma rays, so a new class of DM scenarios with ''leptophilic'' WIMP candidates have been  invoked.
 On the other hand, the increase in the positron fraction, which could have had any spectral shape for
 new physics, matches well the shape expected from proton background.
\end{abstract}

\maketitle


\section{Introduction}

The quest for understanding the nature of the elusive dark matter has received a new boost
by the rising positron fraction observed by the PAMELA space-borne experiment \citep{Adriani:2008zr}
in combination with rather hard spectra of electrons and positrons above expected background by the
FERMI satellite \citep{Abdo:2009zk}, the ATIC balloon experiment \citep{atic:2008zzr}
and the H.E.S.S. earth-bound Cherenkov telescope \citep{Aharonian:2009ah}.
These excesses were not accompanied by an obvious excess in diffuse gamma rays
in the halo at mid-latitude \citep{Porter:2009sg} nor in antiprotons \citep{Adriani:2008zq},
which has led to speculations about a new class of ''leptophilic'' dark matter candidates
\citep{ArkaniHamed:2008qn,Nomura:2008ru}, which fit the data \citep{Bergstrom:2009fa}.
On the other hand astrophysical explanations, like nearby pulsars
\citep{Atoyan:2000rg,Profumo:2008ms,Serpico:2008te,Hooper:2008kg,Blasi:2009hv,Grasso:2009ma,Chowdhury:2009jd,Yuksel:2008rf}
or nearby SNRs \citep{Blasi:2009hv,Fujita:2009wk,Shaviv:2009bu}, provide viable explanations.

In this short summary we first discuss the data and then the possible explanations in terms of
astrophysical sources or dark matter annihilation.  Given the explosion of papers it will not be
possible to discuss all ideas. Recent reviews on dark matter can be found in Refs.
\citep{Bergstrom:2009ib} and \citep{Hooper:2009zm}.

\begin{figure*}
  \includegraphics[width=0.45\textwidth]{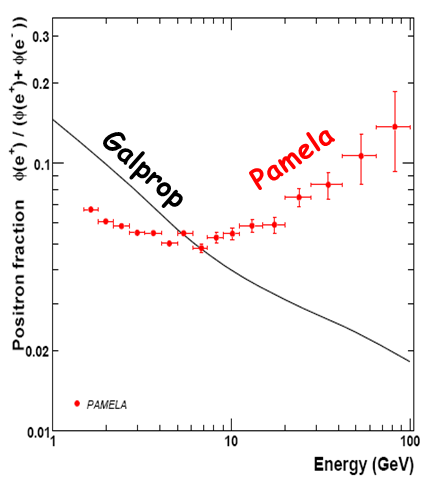}
  \includegraphics[width=0.45\textwidth]{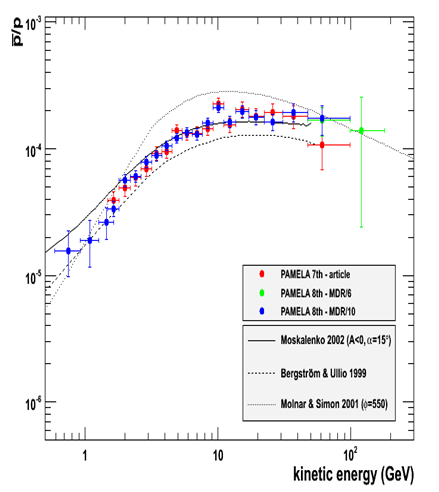}\\
  \caption{Left: Positron fraction as measured by PAMELA in comparison with the GALPROP prediction \citep{Adriani:2008zr}.
  Right: PAMELA antiproton/proton ratio compared with predictions \citep{Adriani:2008zq}.}
\label{f1}
\end{figure*}

\section{PAMELA data on positrons and antiprotons}
PAMELA (Payload for Antimatter Matter Exploration
and Light-nuclei Astrophysics) is a satellite-borne experiment carried on board of a Russian earth observing satellite,
which was launched in June, 2006. Its main components are i) a magnetic spectrometer consisting
of a silicon tracker inside a  45 cm high permanent magnet with a square inner bore of 16x13 cm$^2$
and a field of 0.4T; ii) a 24x24 cm$^2$ Si-tungsten calorimeter with a thickness of 16.3 X$_0$;
iii) a time-of-flight counter iv) an anti-coincidence counter; v) a neutron counter.
The main purpose of the experiment is the measurement of the antiproton and positron components
of cosmic rays in a energy range of 50 MeV to a few 100 GeV  with high statistics.

The main surprise of the first two years of data taking was the observation of
 an increasing positron fraction, defined as the fraction $e^+/e^+ +e^-$, above  10 GeV, while
 standard propagation models, like GALPROP, predict a continuously decreasing fraction
 \citep{Adriani:2008zr}. The data up to 100 GeV are shown in Fig. \ref{f1}a and
 agree with previous measurements up to 10 GeV by HEAT \citep{Barwick:1997ig} and
 AMS-01 \citep{Aguilar:2007yf}.
In contrast, the antiprotons did not show any particular feature, as shown by the antiproton/proton ratio
 in  Fig. \ref{f1}b \citep{Adriani:2008zq}.

\section{Electron plus positrons from ATIC,  FERMI and HESS}
\begin{figure*}
  \includegraphics[width=0.45\textwidth]{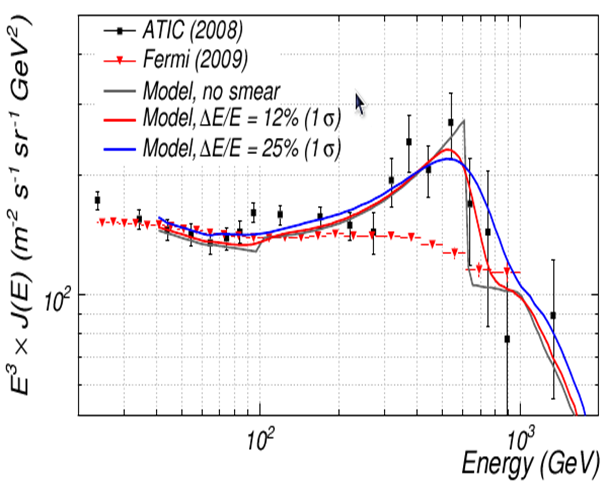}
  \includegraphics[width=0.45\textwidth]{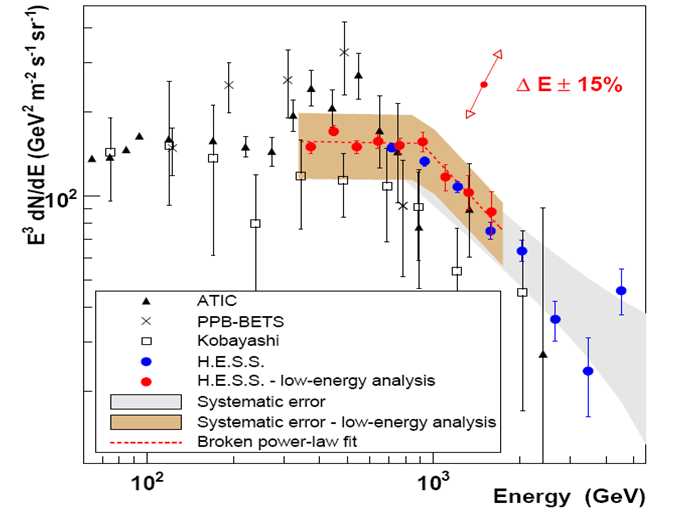}\\
  \caption{Left: Spectrum of electrons+positrons from FERMI and ATIC. From  \citet{Abdo:2009zk}.
  Right: as left, but for different experiments and including new data from H.E.S.S. From \citet{Aharonian:2009ah}.}
\label{f2}
\end{figure*}
The Advanced Thin Ionization Calorimeter (ATIC) is a balloon-based experiment designed to
study the spectra of cosmic ray protons, light nuclei, and electrons.
It measures electromagnetic showers in a 16 X$_0$ thick calorimeter. The spectrum of the summed fluxes of positrons
and electrons start to become harder
above 300 GeV with an abrupt decrease at around 600 GeV, as shown in Fig. \ref{f2}a \citep{atic:2008zzr}.
The rather steep decline, observed in 3 independent flights, has led to  speculations about new
physics, especially the possibility of the dark matter annihilation of Kaluza-Klein type WIMPs.
These particles occur naturally in theories of universal extra dimensions as reviewed by
\citet{Hooper:2007qk}. They can decay into lepton pairs, which would lead to a peak in the spectrum
smeared with a low energy tail from final state radiation. Given that the excess stops around 800 GeV
this would require WIMP masses in this range.
However, the bump was not confirmed by data from the FERMI telescope \cite{Abdo:2009zk}, a gamma-ray telescope
launched in June 2008 by the NASA. It consists of a silicon tracker including a converter to convert gamma rays
into electron pairs, whose energy is measured in a CsI(TI) scintillating calorimeter with a thickness of  8.6 X$_0$.
The FERMI data showed a smooth spectrum with only a shallow structure, as shown by the dashed red line
in Fig. \ref{f2}a.  The  spectrum up to 1 TeV could be reasonably well described by a power line
proportional to $E^{-3.0}$ in agreement with data from the H.E.S.S. experiment
(High Energy Stereoscopic System, an earth-bound array of Cherenkov telescopes,
which measures  high energy gamma rays
 from the  Cherenkov light in the air shower), as shown in Fig. \ref{f2}b.
 Also these   data
show  no indication of a structure in the electron spectrum, but
rather a power-law spectrum with a spectral index of $3.0 \pm 0.1(stat.)\pm 0.3(syst.)$, which steepens at
about 1 TeV \citep{Aharonian:2009ah}, indicative of the cut-off of nearby astrophysical sources.
\section{Constraints from Antiprotons}
Charged particles frequently change directions in the
irregular magnetic fields of the Galaxy. This kind of random walk is usually described by a diffusion
equation.
The diffusion equation describing the transport of cosmic rays (CRs) in the Galaxy is
solved numerically in the publicly available GALPROP code \citep{GalpropMan}.
In its simplest version, sometimes called the conventional model,
the diffusion coefficient is isotropic and has everywhere the same  value and
the spectral shape of the locally observed CRs is assumed
to be representative for the Galaxy.
Details about Galactic propagation models can be found in a
review by \citet{Strong:2007nh}.

\begin{figure*}
  \includegraphics[width=0.48\textwidth]{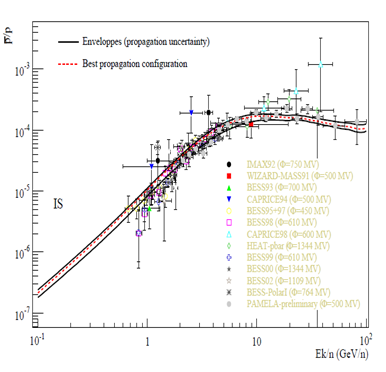}
  \includegraphics[width=0.45\textwidth]{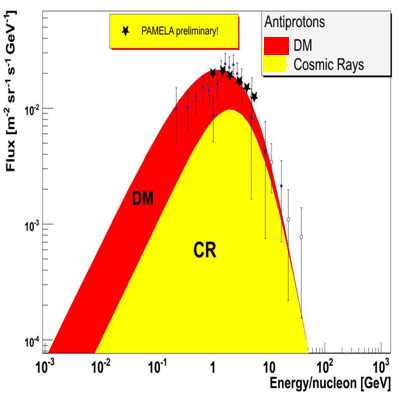}\\
  \caption{Left: Antiproton/proton ratio in an isotropic propagation model from \citet{Donato:2008jk}.
  Right: Antiproton spectrum in an anisotropic propagation model.}
\label{f3}
\end{figure*}

The isotropic propagation models have a few important deficiencies: i) they do not allow for significant convection
of CRs by gas streams, which are therefore assumed to have negligible velocities. However,
this  disagrees with the recent ROSAT data on X-ray spectra in the halo \citep{Breitschwerdt:2008na}
ii) they cannot explain the large bulge/disc ratio of the 511 keV
positron annihilation line, as observed by INTEGRAL, see Ref. \citep{Prantzos:2008rn} for a review
iii) they tend to overproduce the number of gamma rays
towards the Galactic centre in comparison to the gamma ray production away from the Galactic centre
\citep{Breitschwerdt:2002vs}.
These deficiencies can be remedied if one allows for a different diffusion constant in the halo and
the disc \citep{Gebauer:2009hk}. In such a model the observed amount of antiprotons tends to be a factor of
about two below the observed rate. Such a tendency was already observed before for the conventional
GALPROP model \citep{Moskalenko:1998id} and disagrees with semi-analytical calculations, which can describe
the observed antiproton fluxes reasonably well, as shown in Fig. \ref{f3}a.
The antiproton production in an anisotropic propagation
model is shown in Fig. \ref{f3}b. The secondary antiproton production is roughly a factor two below the
data, so one needs an additional source of antiprotons in such a model. This can be nicely provided
by the annihilation of a  60 GeV neutralino, as shown by the red area. Note that low mass WIMPs have
a spectral shape similar to the secondary (+tertiary) antiprotons, so their contributions cannot be
distinguished from the spectral shape. In such models half of the antiprotons can be attributed to dark matter
annihilation, thus relaxing the constraints on DMA models discussed by \citet{Donato:2008jk}.

\begin{figure*}
  \includegraphics[width=0.8\textwidth]{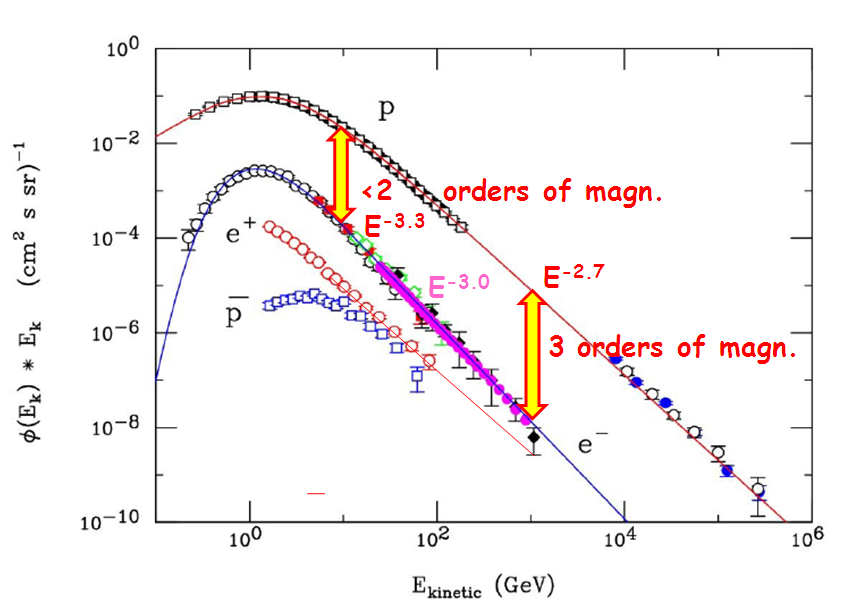}
  \caption{CR spectra. Adapted from \citet{lipari}.}
\label{f4}
\end{figure*}

\section{Interpretation}

The data described in the previous sections has caused an explosion of papers on possible interpretations
in terms of new physics, varying from new astrophysical sources to dark matter annihilation.
Unfortunately, none of the explanations can be excluded and what is worse, they all seem possible,
so they can contribute simultaneously.
For astrophysical sources pulsars are outstanding candidates for positrons,
but also secondary acceleration of positrons in SNR may contribute or local nearby
sources with correspondingly lower energy losses may harden the positron fraction.
Also hadronic background may be non-negligible, so the total contribution to the
positron fraction may be described by
\begin{equation}\label{ep}
    |Data>= a_{bg}|Backgr>+
    a_{pls}|Pulsar>+a_{snr}|SNR>+a_{loc}|Src>+a_{dma}|DMA>
\end{equation}
Of course the sum of the contributions cannot be more than the data, so the coefficients $a$
in front of the contributions are constrained by unitarity.
However, at present it looks like most contribution may saturate the data or
not contribute at all, so most  of the coefficients in Eq. \ref{ep} are allowed
to have values between 0 and 1. Clearly, more independent data are needed to
constrain the different contributions. We will now discuss in more detail the various contributions.
\subsection{Background} \label{uncertainties}
Separating antimatter from matter particles in cosmic rays (CRs) is not an easy task,
because of the preponderance of protons, followed by electrons, which have already a flux,
which is several orders of magnitude lower, as  demonstrated in Fig. \ref{f4} \citep{lipari}.

\begin{figure*}
  \includegraphics[width=0.45\textwidth]{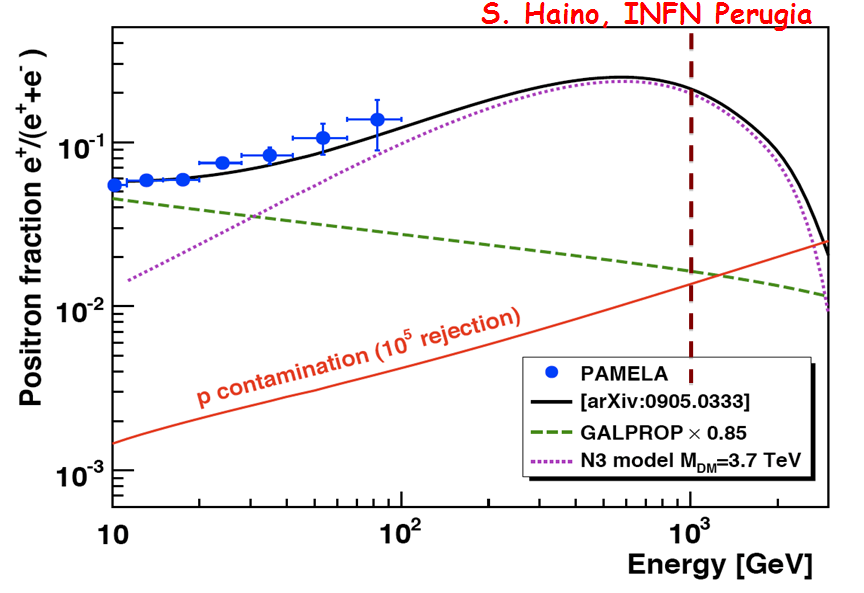}
  \includegraphics[width=0.45\textwidth]{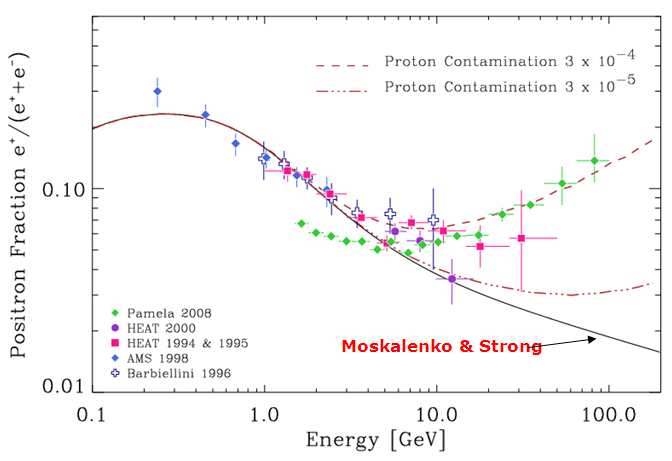}\\
  \caption{Left: AMS-02 hadronic background to the positron fraction for a proton rejection level
  of 10$^{-5}$.
  Note that the indicated DM model to describe the PAMELA data points has a similar slope as the hadronic
  background. From \citet{battiston}.
  Right: the same conclusion is reached by the HEAT experiment: if the rejection is 3$\cdot10^{-4}$,
   the positron fraction has  the shape expected from the background. From \citet{tarle}. }
\label{f5}
\end{figure*}

To separate positrons from protons one needs a rejection of at least $1:10^5$, i.e. only one proton
in $10^5$ may simulate a positron, if one wants a background at the percent level. This can be achieved
by a combination of a magnetic spectrometer measuring the charge, a calorimeter for measuring the shower
profile and a transition radiation detector for measuring the velocity. This is the classical design used
in the HEAT experiment and  foreseen for the upcoming AMS-02 experiment.
For PAMELA the same configuration was planned, but  the TRD detector
had to be abandoned in the last moment for technical reasons. Nevertheless, the fine granularity Si-W calorimeter
provides typically a rejection of $10^{-4}$ at 10 GeV to $10^{-5}$ at 100 GeV, as shown by the PAMELA beam
test \citep{Boezio:2006je}.

The only doubt coming to mind about the amount of background arises from the fact,
that for new physics the increase  could
have had any slope, but the increase is exactly what one would expect from an
hadronic background, as shown by the estimates from AMS-02 and HEAT in Fig. \ref{f5}. Therefore
a check of the hadron rejection with an independent detector, like the TRD, would have been useful
and one is looking forward to independent data by the AMS-02 experiment in 2010, which is scheduled
 to be flown to the ISS in the middle of 2010 according to the NASA program.

  The FERMI background rejection is an order of magnitude worse than for PAMELA, because it does not have a
 magnetic spectrometer to measure  charge and momentum. The estimated rejection power is almost flat at a
 level of  $10^{-3}$ below 200 GeV and improves to $10^{-4}$ at 1 TeV \citep{Abdo:2009zk}.
 The hadron rejection therefore needs hard cuts and relies heavily on MC simulation,
 but even if there is a residual hadronic background,
 the wide structure observed by ATIC should  have been observed. In addition, H.E.S.S. did not observe
 a structure (Fig. \ref{f2}b). So the structure seems improbable, but a confirmation of the hard
 electron spectrum from FERMI by the PAMELA data in the region of overlap  would be nice.

\subsection{Pulsars}
\begin{figure*}
  \includegraphics[width=0.5\textwidth]{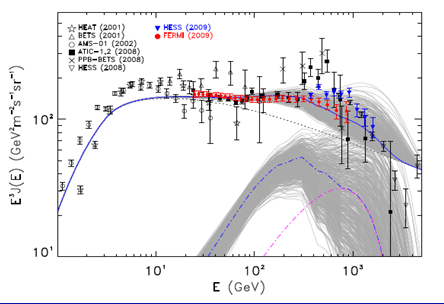}
  \includegraphics[width=0.45\textwidth]{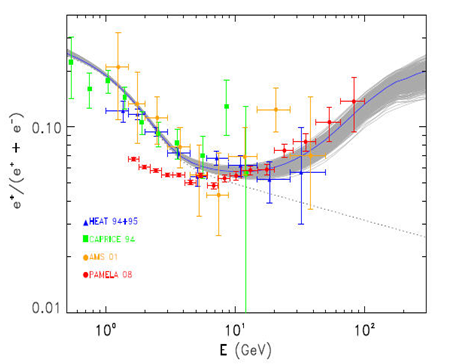}\\
  \caption{Data compared with pulsar model. Left: e$^+$+e$^-$ spectra. Right: positron fraction.
  From Ref. \citep{Grasso:2009ma}.}
\label{f6}
\end{figure*}

Pulsars are rapidly spinning, magnetized
neutron stars, which emit  electromagnetic radiation along the magnetic poles, as can be observed as flashes of
light on Earth, when the poles are directed towards us. They
 are also  prime candidates for sources of primary electrons and positrons, since the strong rotating magnetic
 fields generate photons by synchrotron radiation, which in turn generate electrons and positrons by interaction
 with the magnetic field and these electrons and positrons generate again photons by synchrotron radiation.
 So such a chain reaction delivers lots of positrons and electrons, but no antiprotons. A review about
 pulsars and how they can explain the data was given by \citet{Profumo:2008ms}. The main question is: how
 many charged particles can escape from the strong magnetic fields. This is unknown, since especially young
 pulsars are still surrounded by their nebulae, which trap the electrons and positrons by the magnetic
 fields and energy losses for about   $10^5$ years. Therefore one has to consider pulsars older than this.
Assuming the escape rate and/or the number of old and nearby pulsars to be  free parameters such
pulsars can nicely explain the PAMELA and FERMI data, as shown in Fig. \ref{f6} \citep{Grasso:2009ma}.
If only a few nearby sources contribute,
one could try to search for an anisotropy in the arrival direction of the positrons.
However, even with only a few sources the diffusive scattering washes out anisotropies efficiently.
\subsection{Positron acceleration ins SNRs}
\begin{figure*}
  \includegraphics[width=0.45\textwidth]{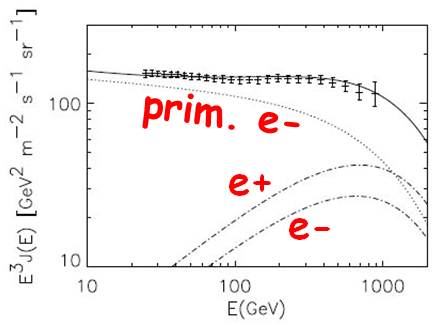}
  \includegraphics[width=0.45\textwidth]{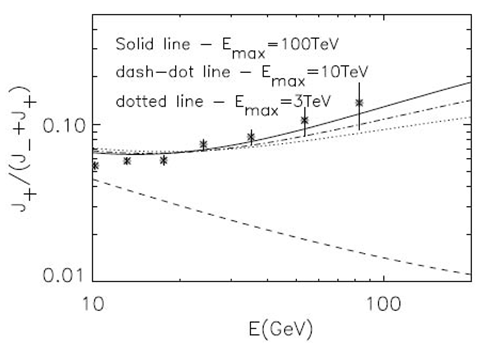}\\
  \caption{Left: electron/positron spectra assuming secondary acceleration in SNR. From Ref. \citep{Blasi:2009hv}.}
\label{f7}
\end{figure*}

The increase in the positron fraction cannot be explained by current sources and propagation models,
since one always expects a decreasing fraction, as shown by the GALPROP curve in Fig. \ref{f1}a and
discussed more generally e.g. by \citet{Serpico:2008te}.
It was suggested by \citet{Blasi:2009hv} that an additional source of positrons could be the secondary
positrons generated and accelerated in old SNRs, especially if they are located inside the high density of
molecular clouds, the birthplaces of stars. If located at a distance of 1-2 kpc, this mechanism would in
addition explain the cut-off at about 1 TeV in agreement with the HESS data \citep{Aharonian:2009ah},
since sources further away would not have such a  hard spectrum due to energy losses. This physical idea
can nicely explain the data on positrons and electrons, as shown in Fig. \ref{f7}, although the strength
of the effect depends heavily on the evolution of SNRs, which is not well known.
Of course, explaining the rise in the positron fraction by acceleration of secondaries would also
increase the antiproton/proton fraction, in disagreement with the data shown in Fig. \ref{f1}b.
However, the increase could happen above 100 GeV, as shown by the simple modeling of Ref. \citep{Blasi:2009bd}.
\subsection{Local Sources}
As mentioned above the electron spectra from local sources might be different from sources further away
due to energy losses. It is clear that in the disc the sources are not homogeneously distributed, since most
of them are in spiral arms. The nearest arm is about 1-2 kpc from the solar neighborhood, so this high density
of sources relative far away will deliver a large fraction of electrons with a relatively soft spectrum. Sources
in the disc  and local sources will deliver increasingly harder spectra and the superposition of all three
components can easily describe the FERMI electron spectrum, as shown in Fig.
\ref{f9}a \citep{Shaviv:2009bu}. Note that the positrons are produced by the CRs in the gas,
so they come from all distances and have a smooth distribution with an intermediately hard spectrum.
The resulting positron fraction in this model first increases and then decreases, similarly to
what would be expected from dark matter annihilation, but different from most other astrophysical explanations.
Clearly, data above 100 GeV will be very interesting to see if it decreases.

 \subsection{Dark Matter Annihilation}
\begin{figure*}
    \includegraphics[width=0.4\textwidth]{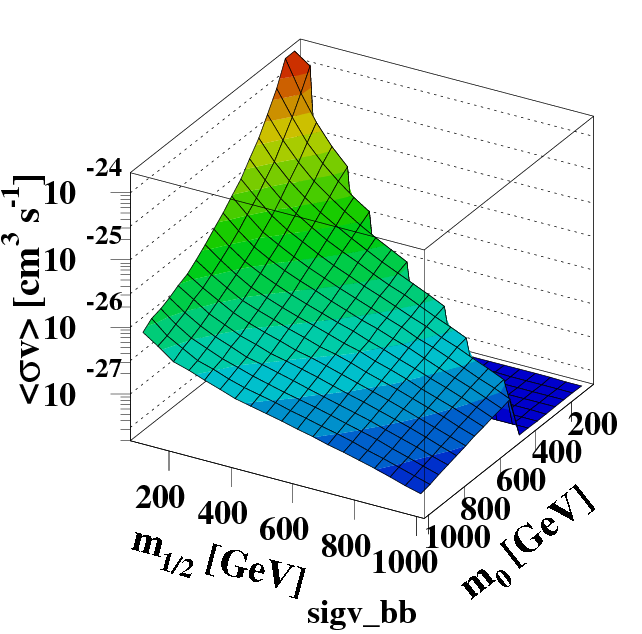}
    \includegraphics[width=0.4\textwidth]{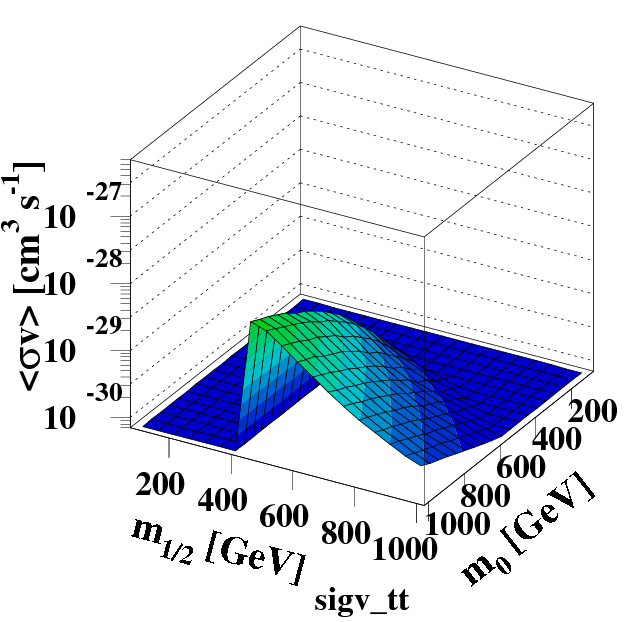}
    \caption{The thermally averaged neutralino
    annihilation  cross section $<\sigma v>$  in the $m_0 m_{1/2}$ plane for $b\overline{b}$ and
$t\overline{t}$ final states.}
    \label{f8}
\end{figure*}
\begin{figure*}
    \includegraphics[width=0.4\textwidth]{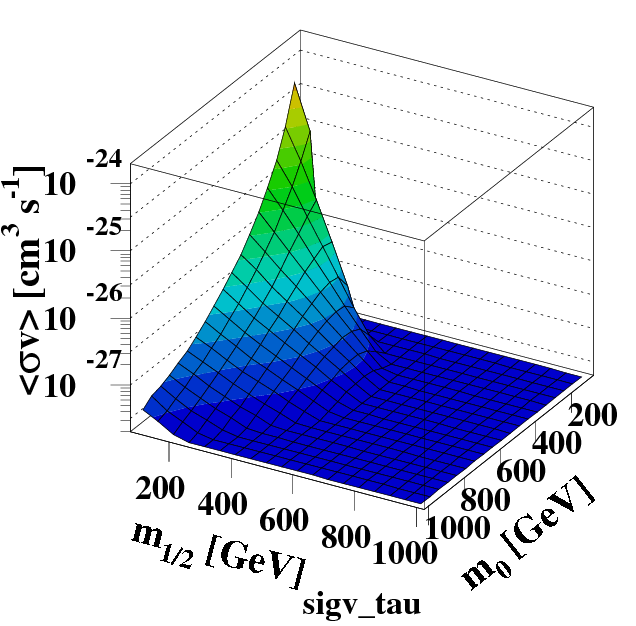}
    \includegraphics[width=0.4\textwidth]{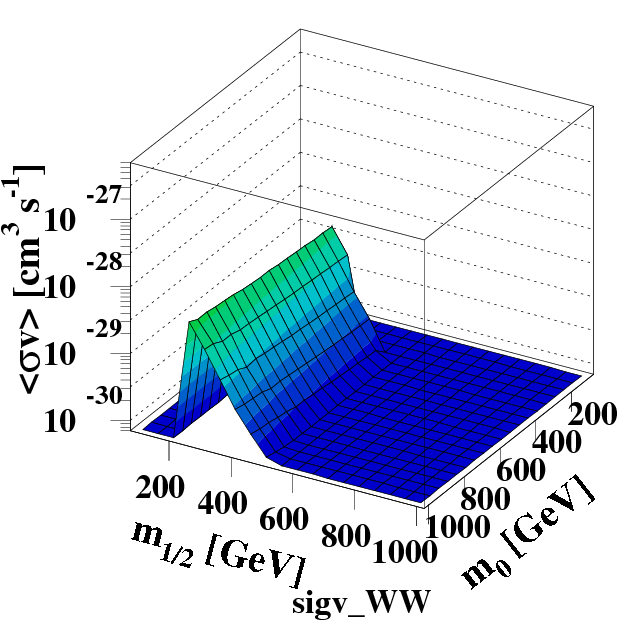}
    \caption{As Fig. \ref{f8}, but for
$t\overline{t}$, $\tau\overline{\tau}$ and $W^+W^-$) final states. }
    \label{f8a}
\end{figure*}
 Dark Matter, if a thermal relic, {\it must} have annihilated or decayed in order to reduce the
 high number density in the early universe, where all particles had similar number densities.
 Nowadays the number density of baryons is 10 orders of magnitude below the photon number density,
 as measured by the temperature of the cosmic microwave background. Given that the DM density,
 given by the WIMP mass times the number density, is at most a factor 6 above the baryonic density,
 the number density of WIMPs must also be  small compared with the photon density, hence the WIMPS
 must have annihilated or decayed, if their number density was similar to the photon density in the
 early universe.

 The lightest neutralino in Supersymmetry is a perfect WIMP candidate, not only because it is
 expected to be stable because of R-parity, but also its annihilation cross section is exactly
 in the right ballpark needed for the 23\% leftover of DM energy in the universe.
 It annihilated preferentially into heavy quarks, like b-quarks, because the annihilation into
 lighter fermions is helicity suppressed, as can be seen  from Figs. \ref{f8} and \ref{f8a}.

The quarks fragment into hadrons with a well-known spectrum and fraction of antiprotons, positrons,
neutrinos and photons, as measured e.g. at the LEP accelerator \citep{Amsler:2008zzb}, so if one observes
a rise in the positron fraction, one expects also an increase in antiproton fluxes. Since the observed
rise in the PAMELA positron fraction is not accompanied by a strong rise in antiprotons, neutralino
annihilation is not a candidate for this rise, at least if one assumes a thermal history of the universe.
In  non-thermal universes dark matter can arise from the decay
of heavy particles, such as the moduli  present in  string theories. In this case freeze-out occurs
at much higher temperatures, i.e. at higher densities, so one needs a  higher annihilation cross section
to get the low relic density observed today \citep{Kane:2009if}. In this case one does not need a large
boost factor. The problem with the antiproton problem stays, so one has to invoke propagation
uncertainties or assume that other sources contribute to the positron flux as well, as discussed before.

\begin{figure*}
  \includegraphics[width=0.4\textwidth]{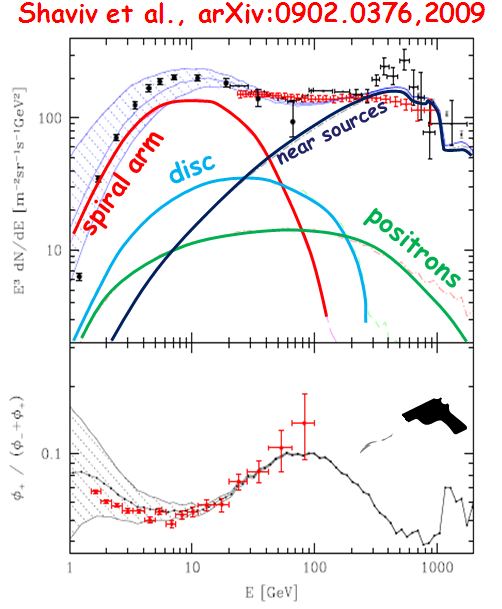}
  \includegraphics[width=0.4\textwidth]{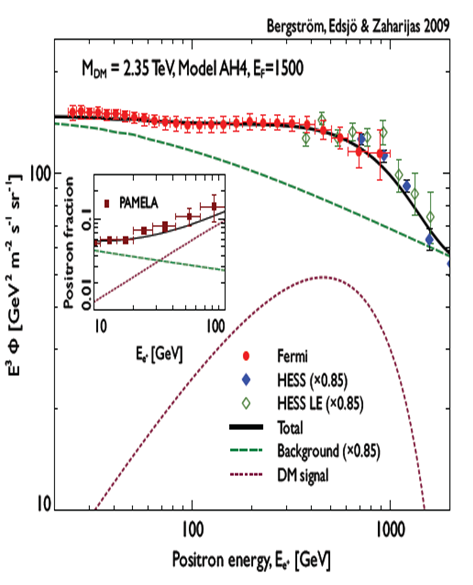}\\
  \caption{Left: Fit of
  astrophysical sources to the FERMI electron spectrum and
  the PAMELA positron fraction. From Ref. \citep{Shaviv:2009bu}.
  Right: Fit of  a leptophilic DM candidate to the FERMI electron spectrum and
  the PAMELA positron fraction. From Ref. \citep{Bergstrom:2009fa}.  }
\label{f9}
\end{figure*}

If one ignores astrophysical contributions and investigates the possibility that DMA is completely
responsible for the leptonic excesses then one has to resort to leptophilic WIMP candidates.
This can happen if the WIMPS decay into new states, which are light enough
to prevent antiprotons in the final states, i.e. typical WIMP masses are below 1 GeV.
For example,
\citet{Nomura:2008ru} propose DM fermions decaying into an axion and a scalar with the latter decaying
again to axions. For an axion  mass in the range 360 - 800 MeV antiproton production is forbidden.
Alternatively, \citet{ArkaniHamed:2008qn} propose a new force carrier $\phi$ of a few GeV,
e.g. a pseudo-Goldstone
boson, into which a heavy WIMP can annihilate.
Resumming the ladder diagrams of the $\phi$ exchange can enhance
the annihilation cross section for non-relativistic particles, i.e. WIMPs in the TeV range, by
orders of magnitude \citep{Hisano:2004ds}, thus providing an explanation for the large cross section needed
to explain the PAMELA and FERMI excesses via this so-called Sommerfeld enhancement \citep{Iengo:2009ni}.
Such models could  also help explaining  the large bulge/disc ratio of the INTEGRAL positron annihilation
line \citep{Prantzos:2008rn}. However, as in the case for the
PAMELA and FERMI excesses, a
very natural astrophysical explanation exists: Galactic winds, as observed by ROSAT,
convect the positrons from the disc into the halo, thus suppressing the positron annihilation in the disc
\citep{Gebauer:2009hk}.
Nevertheless,  independent of astrophysical explanations, models with new force carriers are able
to explain nicely the excesses in positrons and electrons as well,
as shown in Fig. \ref{f9}b \citep{Bergstrom:2009fa}.

\section{Conclusion}
\begin{figure*}
  \includegraphics[width=0.75\textwidth]{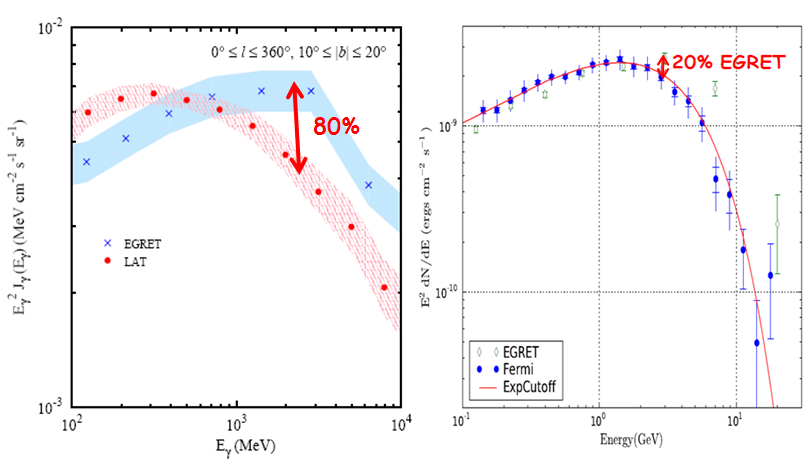}
  \caption{Left: preliminary FERMI data on diffuse gamma rays compared with EGRET data in
  the mid-latitude range ($10^0 <b<20^0$) \citep{Porter:2009sg}.
  Right: published FERMI data from VELA pulsar compared with EGRET data \citep{Abdo:2008ef}.
  Clearly, the published FERMI data agrees better with the EGRET data than the preliminary diffuse gamma ray data
  in the region around a few GeV.}
\label{f10}
\end{figure*}
The observed excesses in electrons and positrons by PAMELA, FERMI, HESS and ATIC have led to an interesting
mix of proposals, varying from new astrophysical sources to DM annihilation. Unfortunately, all seem possible,
which indicates how difficult it is to extract  conclusive evidence about the origin of these charged CRs.
This difficulty is intimately related to the fact that charged particles do not point to their origin
but change direction by the Galactic magnetic fields. Gamma rays do point to the origin, so point
sources can be subtracted and the isotropic extragalactic contribution can be estimated as well.

FERMI will hopefully  soon publish their data on diffuse gamma rays with energies ranging from
tens of MeV to hundreds of GeV, i.e. the range of interest for typical WIMP scenarios;
preliminary data indicate that the GeV excess
in the diffuse gamma rays seen by
EGRET is not observed anymore \citep{Porter:2009sg}, although the calibration via the VELA pulsar
did not show a large discrepancy between EGRET and FERMI in the region of the GeV excess \citep{Abdo:2008ef},
as shown in Fig. \ref{f10}. Of course, the background for the diffuse gamma ray data is larger than
for the strong VELA pulsar, which may indicate that the background was not under control in the EGRET experiment,
since with the higher spatial resolution of the FERMI tracker and the better MC simulation
the background should be better under control in the FERMI data.

Clearly more independent data are welcome, which hopefully will be provided with the AMS-02 spectrometer to
be flown to the ISS in the middle of 2010. But above all it is hoped that the LHC will provide direct clues
to the nature of DM by finding WIMP candidates, which can then be compared to the indirect and direct searches
for DM.\\[2mm]
{\bf Acknowledgments.}
I wish to thank Pran Nath for organizing an excellent SUSY09 conference and many colleagues
for illuminating discussions. This work was supported by the Deutsches Zentrum f\"ur Luft- und Raumfahrt (DLR).


\begin{thebibliography}{42}
\expandafter\ifx\csname natexlab\endcsname\relax\def\natexlab#1{#1}\fi
\providecommand{\enquote}[1]{``#1''}
\expandafter\ifx\csname url\endcsname\relax
  \def\url#1{\texttt{#1}}\fi
\expandafter\ifx\csname urlprefix\endcsname\relax\def\urlprefix{URL }\fi
\providecommand{\eprint}[2][]{\url{#2}}

\bibitem[Adriani et~al.(2009{\natexlab{a}})]{Adriani:2008zr}
O.~Adriani, et~al., \emph{Nature} \textbf{458}, 607--609 (2009{\natexlab{a}}),
  \eprint{0810.4995}.

\bibitem[Abdo et~al.(2009{\natexlab{a}})]{Abdo:2009zk}
A.~A. Abdo, et~al., \emph{Phys. Rev. Lett.} \textbf{102}, 181101
  (2009{\natexlab{a}}), \eprint{0905.0025}.

\bibitem[Chang et~al.(2008)]{atic:2008zzr}
J.~Chang, et~al., \emph{Nature} \textbf{456}, 362--365 (2008).

\bibitem[Aharonian et~al.(2009)]{Aharonian:2009ah}
F.~Aharonian, et~al.  (2009), \eprint{0905.0105}.

\bibitem[Porter(2009)]{Porter:2009sg}
T.~A. Porter  (2009), \eprint{0907.0294}.

\bibitem[Adriani et~al.(2009{\natexlab{b}})]{Adriani:2008zq}
O.~Adriani, et~al., \emph{Phys. Rev. Lett.} \textbf{102}, 051101
  (2009{\natexlab{b}}), \eprint{0810.4994}.

\bibitem[Arkani-Hamed et~al.(2009)]{ArkaniHamed:2008qn}
N.~Arkani-Hamed, D.~P. Finkbeiner, T.~R. Slatyer, and N.~Weiner, \emph{Phys.
  Rev.} \textbf{D79}, 015014 (2009), \eprint{0810.0713}.

\bibitem[Nomura and Thaler(2009)]{Nomura:2008ru}
Y.~Nomura, and J.~Thaler, \emph{Phys. Rev.} \textbf{D79}, 075008 (2009),
  \eprint{0810.5397}.

\bibitem[Bergstrom et~al.(2009)]{Bergstrom:2009fa}
L.~Bergstrom, J.~Edsjo, and G.~Zaharijas, \emph{Phys. Rev. Lett.} \textbf{103},
  031103 (2009), \eprint{0905.0333}.

\bibitem[Atoyan et~al.(2005)]{Atoyan:2000rg}
A.~M. Atoyan, F.~A. Aharonian, R.~J. Tuffs, and H.~J. Volk, \emph{Astr.\&
  Astroph.} \textbf{294}, L41 (2005).

\bibitem[Profumo(2008)]{Profumo:2008ms}
S.~Profumo  (2008), \eprint{0812.4457}.

\bibitem[Serpico(2009)]{Serpico:2008te}
P.~D. Serpico, \emph{Phys. Rev.} \textbf{D79}, 021302 (2009),
  \eprint{0810.4846}.

\bibitem[Hooper et~al.(2009)]{Hooper:2008kg}
D.~Hooper, P.~Blasi, and P.~D. Serpico, \emph{JCAP} \textbf{0901}, 025 (2009),
  \eprint{0810.1527}.

\bibitem[Blasi(2009)]{Blasi:2009hv}
P.~Blasi, \emph{Phys. Rev. Lett.} \textbf{103}, 051104 (2009),
  \eprint{0903.2794}.

\bibitem[Grasso et~al.(2009)]{Grasso:2009ma}
D.~Grasso, et~al., \emph{Astropart. Phys.} \textbf{32}, 140--151 (2009),
  \eprint{0905.0636}.

\bibitem[Chowdhury et~al.(2009)]{Chowdhury:2009jd}
D.~Chowdhury, C.~J. Jog, and S.~K. Vempati  (2009), \eprint{0909.1182}.

\bibitem[Yuksel et~al.(2009)]{Yuksel:2008rf}
H.~Yuksel, M.~D. Kistler, and T.~Stanev, \emph{Phys. Rev. Lett.} \textbf{103},
  051101 (2009), \eprint{0810.2784}.

\bibitem[Fujita et~al.(2009)]{Fujita:2009wk}
Y.~Fujita, et~al., \emph{Phys. Rev.} \textbf{D80}, 063003 (2009),
  \eprint{0903.5298}.

\bibitem[Shaviv et~al.(2009)]{Shaviv:2009bu}
N.~J. Shaviv, E.~Nakar, and T.~Piran  (2009), \eprint{0902.0376}.

\bibitem[Bergstrom(2009)]{Bergstrom:2009ib}
L.~Bergstrom  (2009), \eprint{0903.4849}.

\bibitem[Hooper(2009)]{Hooper:2009zm}
D.~Hooper  (2009), \eprint{0901.4090}.

\bibitem[Barwick et~al.(1997)]{Barwick:1997ig}
S.~W. Barwick, et~al., \emph{Astrophys. J.} \textbf{482}, L191--L194 (1997),
  \eprint{astro-ph/9703192}.

\bibitem[Aguilar et~al.(2007)]{Aguilar:2007yf}
M.~Aguilar, et~al., \emph{Phys. Lett.} \textbf{B646}, 145--154 (2007),
  \eprint{astro-ph/0703154}.

\bibitem[Hooper and Profumo(2007)]{Hooper:2007qk}
D.~Hooper, and S.~Profumo, \emph{Phys. Rept.} \textbf{453}, 29--115 (2007),
  \eprint{hep-ph/0701197}.

\bibitem[Strong and Moskalenko(2006)]{GalpropMan}
A.~Strong, and I.~Moskalenko,
  \emph{http://galprop.stanford.edu/manuals/manual.pdf}  (2006).

\bibitem[Strong et~al.(2007)]{Strong:2007nh}
A.~W. Strong, I.~V. Moskalenko, and V.~S. Ptuskin, \emph{Ann. Rev. Nucl. Part.
  Sci.} \textbf{57}, 285--327 (2007), \eprint{astro-ph/0701517}.

\bibitem[Donato et~al.(2009)]{Donato:2008jk}
F.~Donato, D.~Maurin, P.~Brun, T.~Delahaye, and P.~Salati, \emph{Phys. Rev.
  Lett.} \textbf{102}, 071301 (2009), \eprint{0810.5292}.

\bibitem[Breitschwerdt(2008)]{Breitschwerdt:2008na}
D.~Breitschwerdt, \emph{Nature} \textbf{452}, 826 (2008).

\bibitem[Prantzos(2008)]{Prantzos:2008rn}
N.~Prantzos, \emph{New Astron. Rev.} \textbf{52}, 457--459 (2008),
  \eprint{0809.2491}.

\bibitem[Breitschwerdt et~al.(2002)]{Breitschwerdt:2002vs}
D.~Breitschwerdt, V.~A. Dogiel, and H.~J. Volk, \emph{Astron. Astrophys.}
  \textbf{385}, 216--238 (2002), \eprint{astro-ph/0201345}.

\bibitem[Gebauer and de~Boer(2009)]{Gebauer:2009hk}
I.~Gebauer, and W.~de~Boer  (2009), \eprint{0910.2027}.

\bibitem[Moskalenko et~al.(1998)]{Moskalenko:1998id}
I.~V. Moskalenko, A.~W. Strong, and O.~Reimer, \emph{Astron. Astrophys.}
  \textbf{338}, L75--L78 (1998), \eprint{astro-ph/9808084}.

\bibitem[Lipari(2009)]{lipari}
P.~Lipari, \emph{Pamela Physics Workshop, Rom}  (2009),
  \eprint{http://pamela.roma2.infn.it/workshop09/slides_WS2009/Lipari.pdf}.

\bibitem[Battiston(2009)]{battiston}
R.~Battiston, \emph{private communication}  (2009).

\bibitem[Tarle(2009)]{tarle}
G.~Tarle, \emph{Talk at PPC09, Oklahoma City}  (2009).

\bibitem[Boezio et~al.(2006)]{Boezio:2006je}
M.~Boezio, et~al., \emph{Astropart. Phys.} \textbf{26}, 111--118 (2006).

\bibitem[Blasi and Serpico(2009)]{Blasi:2009bd}
P.~Blasi, and P.~D. Serpico, \emph{Phys. Rev. Lett.} \textbf{103}, 081103
  (2009), \eprint{0904.0871}.

\bibitem[Amsler et~al.(2008)]{Amsler:2008zzb}
C.~Amsler, et~al., \emph{Phys. Lett.} \textbf{B667}, 1 (2008).

\bibitem[Kane et~al.(2009)]{Kane:2009if}
G.~Kane, R.~Lu, and S.~Watson  (2009), \eprint{0906.4765}.

\bibitem[Hisano et~al.(2005)]{Hisano:2004ds}
J.~Hisano, S.~Matsumoto, M.~M. Nojiri, and O.~Saito, \emph{Phys. Rev.}
  \textbf{D71}, 063528 (2005), \eprint{hep-ph/0412403}.

\bibitem[Iengo(2009)]{Iengo:2009ni}
R.~Iengo, \emph{JHEP} \textbf{05}, 024 (2009), \eprint{0902.0688}.

\bibitem[Abdo et~al.(2009{\natexlab{b}})]{Abdo:2008ef}
A.~A. Abdo, et~al., \emph{Astrophys. J.} \textbf{696}, `1084--1093
  (2009{\natexlab{b}}), \eprint{0812.2960}.

\end{thebibliography}

\end{document}